# Analysis of Increasing Malwares and Cyber Crimes Using Economic Approach

Umer Asgher
SMME
College of E&ME
National University of Science and Technology (NUST)
Islamabad, Pakistan
umer_asgher2000@yahoo.com

Fahad Moazzam Dar
College of E&ME
National University of Science and Technology (NUST)
Islamabad, Pakistan
fahaddar@gmail.com

Ali Hamza
College of E&ME
National University of Science and Technology (NUST)
Islamabad, Pakistan
ahrocs752@gmail.com

Abdul Moeed Paracha
College of E&ME
National University of Science and Technology (NUST)
Islamabad, Pakistan.
moeed.paracha@gmail.com

*ABSTRACT-* *The economics of an internet crime has newly developed into a field of controlling black money. This economic approach not only provides estimated technique of analyzing internet crimes but also gives details to analyzers of system dependability and divergence. This paper will highlight on the subject of online crime, which has formed its industry since. It all started from amateur hackers who cracked websites and wrote malicious software in pursuit of fun or achieving limited objectives to professional hacking. In the past days, electronic fraud was main objective but now it has been changed into electronic hacking. This study focuses the issue through an economic analysis of available web forum to deals in malware and private information. The findings of this survey research provide considerable in-depth sight into the functions of malware economy spinning around computer impositions and compromise. In this regard, the survey research paper may benefit particularly computer security officials, the law enforcement agencies, and in general prospective anyone involved in better understanding cybercrime from the offender standpoint.*

I. Introduction

The internet usage becomes very important each day around the world. Number of Internet users is taking benefit of the convenience and flexibility the internet has given to them, searching from information to online access for e-business and e- finance. This has given birth to the possessions own by the ordinary Internet consumer, malicious attacker find out the Web because a novel field for building money by harming naive internet consumer. Much progress has been made by such persons for the growth of online black markets that facilitate them to sale and distribute tools and information that is intended to negotiation computer networks and consumer. particularly, a number of web medium permit individuals to purchase malicious software to victimize vulnerable systems and persons. Now it is obvious that malicious players can obtain funds to make possible criminal action, however it is not obvious what the revisit on investments is like relative to the expenses of trade goods and services in the course of these market. Due to this reason the past years have witnessed the increase in efforts to increase the security of systems and networks. Hacking is a cybercrime[1],its purpose, procedure adopted, goal and threats involved can be classified as:-

Purpose: self esteem, agenda, moral work, front for other organizations. Procedure: Black money, recruitment of some insiders of organization, malware and hacking that target a system. Goal: Large firms, governments, supply chains managers, security chain or media outlets that oppose their agenda or moral work. Threat: It depends on the nature of protected information and the public potential targets; in the section 2 we shall cover the types of malwares. In section 3, cyber crime motivation. In section 4, we shall be covering analysis of survey research. In section 5 IBM's X-Force compiled report. In section 6 Norton cyber crime report analysis is covered followed by findings of the research and the conclusions.

II. Types of Malware
There are numerous types of malwares including:-

--Smartphone Malware

--Worms, viruses and other propagating Malware

--Spyware, keystroke loggers, information theft Malware

--Botnet attacks, detection/tracking and defense

--Root kit and virtualization techniques

III. 3. Analysis of Survey Research

Cyber activity has the following main motivations:-

--Malicious activity by country against another country

--Web-based attacks on the users

--Data breaches leading to identity theft

--Bot-infected computers

## IV. 4. Analysis of Survey Research

### A. The Online crime Market

The online crime market or the cybercrime market is, as its name suggests, a pool of cybercriminals who seek to make lots of dirty money out of deceiving unknowing web users [3]. It's made up of malicious networks that, although they're already filled with cybercriminals' interaction, they sometimes interact between them as well. Now, who drives it exactly? Well, any cyber-pro with malicious intentions, who wants to breach your internet security. In short, as with any other market, the online black market includes activities like: production (of malware), logistics (delivering malware to web users and catching their online data), sales (of users' data to other cybercriminals), marketing and promotion of cyber-products obtained illegally [2][3].

### B. Mechanism of the Online crime Market

1. Phase 1. Malware creation

Behind every Trojan, virus, worm, bot and other malware there's business. Heads of criminal networks contact programmers to develop malware, hackers to break into networks or other scammers and fraudsters to devise spam and phishing attacks [4]. The victims are usually Windows users and web users tricked while browsing for specific information, banking, socializing. If you fall for one of their internet security scams, your data may get stored onto servers that hackers can access. They can use it to enter your accounts and steal money and your identity, or trade it on the online black market.

2. Phase 2. Promotions of illegal online commodities

Just like legal commodity markets, the online black market is very competitive. And in order to be profitable, cybercriminals have to promote their "goods" to fellow-criminals. They come up with promotions, demos, service guarantees, even discounts for large "purchases" all advertised on underground forums and sometimes on social media.

3. Phase 3. The sales process

A cyber crook gets interested in another cyber crook's offer. Here is how the sales process work:-

--Contact the client/vendor via private chat or e-mail using generic addresses and negotiate.

--Use existing underground online stores to distribute the products.

--Establish the method of payment, which is always one that everybody uses, like Western Union.

--Ask for customer support if the product is not working e.g. if a credit card number is not valid, they will change it for one that is.

4. Phase 4. Money laundering

This phase applies to cyber crooks who steal money from users' bank accounts via bank transfers. As you can imagine this is dirty money that needs to be laundered in order to be used on legal markets. This is where other victims enter the process: money mules attracted by cyber crooks through false job offers. They are promised high commissions just for receiving the stolen money in their bank accounts and then send it to foreign accounts – cyber crooks' accounts. These victims not only get their internet security compromised, but also their physical security.

### C. Working format of internet criminals

Now we design the proposed procedure to explain the interface flanked by different criminals contained by the subversive black work market. Explanation of the economic aspects of the fact and for each criminal; we will highlight their function, what sort of information and merchandise they deal, and what the widespread rate for such goods and services.

1. Virus Creator

Virus creator is malicious Internet users determined by economic earnings and having a very limited of technical Knowledge of computer-networking and programming language skills .They are some times  able to find vulnerabilities in their organization, or they make use of newly public disclosed vulnerabilities and the matching system. In addition, those people comprise the technical skills work force that develop their own malware based on the original vulnerability intelligence and accessible exploit codes. They sell their operational tools and malware for profit, and offer service to their patrons on demand.

2. Website Designers

The second level of persons involved within this market is Website Designers [5]. These persons attract visitors to their website with the assist of free stuff like free online movies, online gaming, free music. These websites then deceive their web visitors and put up for sale the transfer or their load to the Envelopes pirates, by host these the web based Trojans or other viruses, It results the website visitors to be redirect via malicious web sites to new site then assault the victims and systems roughly. If the bother is positive, a portion of virus or malware is install on the victim's appliance. Web site Designer's can also negotiation well recognized but less security safety measure web sites by exploit probability that exist on these sites by means of the line entrée on the attack machines, they then forward the traffic for this web site to one more infected mechanism, i.e., they then sell the traffic of their victim's web site to the proposed equipment of their choice.

3. Envelopes Pirates

Envelope is a term use in the black - market which mean stolen hacked couple of accounts and passwords, i.e account hacking. We are going to use this term all through our paper. Envelope Pirates have extremely less technical acquaintance and normally purchase ready to use Trojan viruses or even malware generator as of Virus Creator or Website Designers [6]. They do this do to initiate a web-based virus network from which they can make new envelopes. These pirates usually merge a internet base Trojan virus in the midst of a conservative Trojan virus for larceny certain envelope and make relation of the generate Trojan to the web sites. They sell the harvest envelopes to the third party which can be called as Non technical persons, which we introduce in the next section.

4. Non Technical Persons

Non Technical Persons don't contain any specific technical data concerning hacking plus making programmers but they have superior knowledge of the subversive black market itself. If we take example of games, they know which on-line game is at present admired and which profit can be sold for a good price. They purchase these envelopes of account and password from the Envelopes Pirates, and log-in to the objective in order to steal valuable practical possessions like equipment.

5. Buyers of Envelopes

Another group of persons within the entire subversive are the Buyers of Envelopes, which could also be but not require to be Non Technical persons. They put in to the movement part of the industry sequence by set up shops on the internet usually found at eBay. The research will show us that the Buyers of Envelopes typically purchase the possessions from the underground subversive market on internet system with a very less price. They then can sell them to group of actors on the open markets, making large amount of profit due to difference of the money flanked by buying a product and further advertising it.

6. Players

Another party within the whole underground are Players. They are always in contact with the Non technical persons and Envelope Buyers. These Players are always on the civic marketplaces, creating a large amount of profit due to difference of the money stuck between buying a product and further selling it.

7. Market Trends

Now the market trends and details of flow of black money are shown In Figure1, we focus the general idea of the interface of the individual inside the subversive black market. The dealing between these persons inside the bionetwork take place in dissimilar location. The business between Envelopes Buyers, Virus creators, and Web site Designers take place in the hidden black market on dissimilar sort of internet systems. This structure also provide a marketplace for Non Technical persons and Envelope Buyers. The movement of possessions is open on the internet. This is owing to this cause that player are needed to discover for weak control on the movement of stolen possessions.

V. IBM's X-Force Compiled Report

Every year, IBM's X-Force gives a report as in figure2 on the current situation of computer security and its results, what were observed throughout the previous year and its expected results for future.

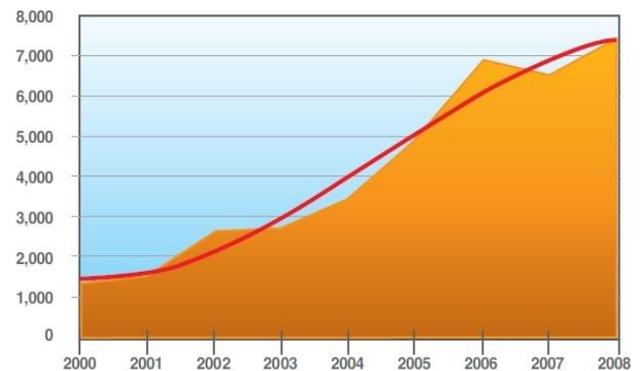

Figure 1. IBM Report

X-Force gave a record tells us that a large number of vulnerability attacks in 2008 there numbers were around 7,406 in total. It gives a total increase of 13.5 percent over 2007. Growth in the Medium and High segments fueled the year increase and more than offset the reduced number of minor flaws. Statistics are shown in figure 3.

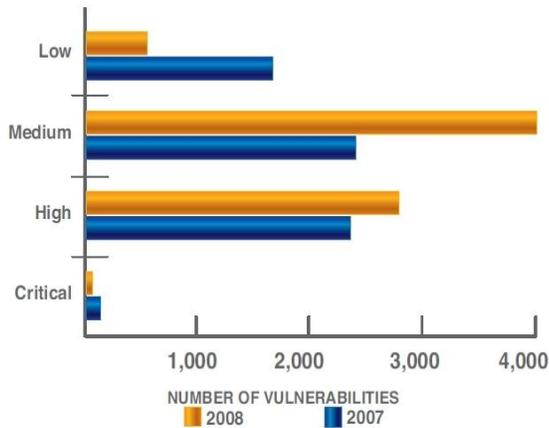

Figure 2. Number of Vulnerabilities

## VI. Norton Cyber Crime Report Analysis

Norton cyber crime report [7] for 2011 suggests 431 million victims worldwide lost $114 billion. Whereas Norton cyber crime report 2012 tells that cyber crime cost 556 Million victims in 24 countries nearly $110 Billion in the year 2012.The Pi graph representation of the year 2012 is shown in figure 4.

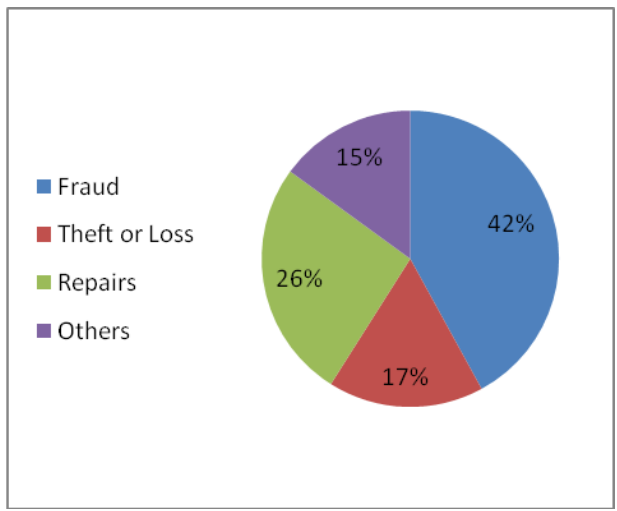

Norton cyber crime report for 2011

The Kaspersky Lab detail [8] for year 2012 is as in figure 5. According to Kaspersky antivirus around 35,000 new pieces of malware and 125,000 new malicious programs appear every single day. Figure 5 clearly shows that main objective of viruses is to deliver and hide malicious program, then to steal data and extort money.

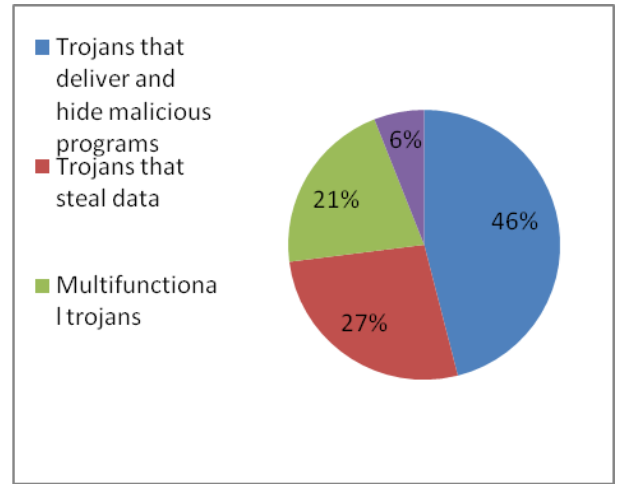

Kaspersky Lab detail for year 2012

## VII. Findings of the Research

The results while getting information from various sources also tell that every minute as many as 232 number computers[9] be contaminated by malware or viruses. The fast pace at which cyber criminals are developing novel malware programs is to make it difficult for worldwide organizations for internet security to manage risk. Most significant defense system can be an intelligence sources and awareness of the potential risks in general public at national and international level. In the coming years the mutual effort by law enforcement agencies [10] and industry to get better information sharing and teamwork all along with the be in motion towards intelligence-driven security will help in giving response to cyber threats in real time.

Government agencies and industries in many countries are already increasing efforts to combat Trojan viruses. Some emerging countries in economy are also experiencing an even greater virus threats due to the less concern from security point of view for the attacks by cybercriminals. Attackers of virus are also relocating their malicious behavior such as phishing hosts, Trojans, bot networks and spam to extra other country from where these attacks can still be directed at targets anywhere in world with less effort from security point of view. However, during year of 2009 a large number of targeted attack occurred that brought these types of incidents into the spotlight of the concerned authorities. The large scale reporting of this attack forced many firms to examine their security infrastructure. Financially motivated attacks by such persons against both firms and individuals remain a great part of the threat matrix around us. [11].

## VIII. Conclusions

To wind up, malware's main objective is to deliver and hide malicious program, then to steal data and extort money. Now this black market has flourished that much that many cybercriminals are creating kits they can sell to new incomers in the underground economy of black market money. It helps all inexperienced new virus attackers with less technical knowledge to cause attacks without too much problem. It may continue to process itself in 2013 and even beyond to keep on one step further of the ever-changing approaches, which businesses use technology. Considering the factors interface of the individual persons and the data analysis of IBM &Norton cyber crime report , at hand are two methods which could be addressed: first of all technologically and second way by educating staff to the risks and best-practice required staying away from these cyber crimes.

In the years to come we may see gradually more and multifaceted malicious software which are intended purposely to be stealthy, something which enterprise must to be extremely conscious of. Making staff aware of the rising hazards accessible by smart phones and further portable digital devices. Since these are in use by more and more people, cyber-criminals will take advantage of this tendency and the linked smugness. People must comprehend that these devices are highly complicated and capable of having malicious code. Technologically, beside the usual updates and malware scans, system administrators ought to run usual anti-root kit sweep as normal practice. With a conventional antivirus way out may not be sufficient to guard against a number of of the most recent forms of malware, so a layered approach in blend with dedicated anti-malware solutions would be necessary and important. Thus beneficiaries of this study are industry, computer security officials, law enforcement agencies, and in general anyone interested in better understanding cybercrime from the offender point of view.

# References

[1] M. Howard, J. Pincus, and J. Wing, "Measuring relative attack surfaces," *Comp. Sec. in the 21st Century*, pp. 109-137, 2005.

[2] P. K. Manadhata and J. M. Wing, "An attack surface metric," *TSE*, vol. 37, pp. 371-386, 2011.

[3] *Symantec Global Internet Security Threat Report: Trends for 2009*, vol. 15, April 2010.

[4] L. Wang, A. Singhal, and S. Jajodia, "Measuring the overall security of network configurations using attack graphs," in *Proc. of DAS'07*, 2007, pp. 98-112.

[5] P. Mell and K. Scarfone, *A Complete Guide to the Common Vulnerability Scoring System Version 2.0. CMU*, 2007.

[6] A. Ozment, "Improving vulnerability discovery models," in *Proc. of QoP'07*, ser. QoP '07. New York, NY, USA: ACM, 2007, pp. 6-11.

[7] *Norton Cybercrime Report*, 2012.

[8] Kaspersky Lab Antivirus Protection Insternet Security, [Online], Retrieved on October 29, 2012, Available: http://www.kaspersky.com/…

[9] A. Arora, R. Krishnan, A. Nandkumar, R. Telang, and Y. Yang, "Impact of vulnerability disclosure and patch availability-an empirical analysis," in *Proc. of WEIS'04*, 2004.

[10] S. Ransbotham and S. Mitra, "Choice and chance: A conceptual model of paths to information security compromise," *ISR*, vol. 20, 2009.

[11] Joel Hruska, *IBM: Malware Economics, Web Security Biggest Issues of 2008,* [Online], Retrieved on October 14, 2012, Available: http://arstechnica.com/security/2009/02/malware-economics-web-security-major-issues-in-2008